\documentclass[twocolumn,twocolappendix]{aastex6}
\usepackage[utf8]{inputenc}
\usepackage{amsmath}
\usepackage{CJKutf8}

\usepackage[normalem]{ulem}
\usepackage{color}
\usepackage{graphicx}
\definecolor{red}{rgb}{1.0,0.0,0.0}

\newcommand{\betapicb}{$\beta$~Pic~b}

\slugcomment{Draft}

\shorttitle{$\beta$ Pic $\mathrm{b}$ Orbit and Transit Prospects} 
\shortauthors{Wang et al.}

\begin{document}
\title{The Orbit and Transit Prospects for $\beta$ Pictoris \lowercase{b} constrained with One Milliarcsecond Astrometry}

\author{
Jason J. Wang\altaffilmark{1},
James R. Graham\altaffilmark{1}, 
Laurent Pueyo\altaffilmark{2}, 
Paul Kalas\altaffilmark{1}, 
Maxwell A. Millar-Blanchaer\altaffilmark{3}, 
Jean-Baptiste Ruffio\altaffilmark{4}, 
Robert J. De Rosa\altaffilmark{1},
S. Mark Ammons\altaffilmark{5},
Pauline Arriaga\altaffilmark{6},
Vanessa P. Bailey\altaffilmark{4},
Travis S. Barman\altaffilmark{7},
Joanna Bulger\altaffilmark{8},
Adam S. Burrows\altaffilmark{9},
Andrew Cardwell\altaffilmark{10,11}, 
Christine H. Chen\altaffilmark{2},
Jeffrey K. Chilcote\altaffilmark{12},
Tara Cotten\altaffilmark{13},
Michael P. Fitzgerald\altaffilmark{6},
Katherine B. Follette\altaffilmark{4},
Ren\'e Doyon\altaffilmark{14},
Gaspard Duch\^ene\altaffilmark{1,15},
Alexandra Z. Greenbaum\altaffilmark{16,2},
Pascale Hibon\altaffilmark{17}, 
Li-Wei Hung\altaffilmark{6},
Patrick Ingraham\altaffilmark{18},
Quinn M. Konopacky\altaffilmark{19},
James E. Larkin\altaffilmark{6},
Bruce Macintosh\altaffilmark{4},
J\'{e}r\^{o}me Maire\altaffilmark{12},
Franck Marchis\altaffilmark{20},
Mark S. Marley\altaffilmark{21},
Christian Marois\altaffilmark{22,23},
Stanimir Metchev\altaffilmark{24,25},
Eric L. Nielsen\altaffilmark{20,4},
Rebecca Oppenheimer\altaffilmark{26},
David W. Palmer\altaffilmark{5},
Rahul Patel\altaffilmark{27},
Jenny Patience\altaffilmark{28},
Marshall D. Perrin\altaffilmark{2},
Lisa A. Poyneer\altaffilmark{5},
Abhijith Rajan\altaffilmark{28},
Julien Rameau\altaffilmark{14},
Fredrik T. Rantakyr\"o\altaffilmark{9},
Dmitry Savransky\altaffilmark{29},
Anand Sivaramakrishnan\altaffilmark{2},
Inseok Song\altaffilmark{13},
Remi Soummer\altaffilmark{2},
Sandrine Thomas\altaffilmark{18},
Gautam Vasisht\altaffilmark{30},
David Vega\altaffilmark{20},
J. Kent Wallace\altaffilmark{30},
Kimberly Ward-Duong\altaffilmark{28},
Sloane J. Wiktorowicz\altaffilmark{31},
and Schuyler G. Wolff\altaffilmark{16,2}
}

\altaffiltext{1}{Astronomy Department, University of California, Berkeley, Berkeley, CA 94720, USA}
\altaffiltext{2}{Space Telescope Science Institute, 3700 San Martin Drive, Baltimore MD 21218 USA}
\altaffiltext{3}{Department of Astronomy \& Astrophysics, University of Toronto, Toronto ON M5S 3H4, Canada}
\altaffiltext{4}{Kavli Institute for Particle Astrophysics and Cosmology, Stanford University, Stanford, CA 94305, USA}
\altaffiltext{5}{Lawrence Livermore National Lab, 7000 East Ave., Livermore, CA 94551, USA}
\altaffiltext{6}{Department of Physics and Astronomy, UCLA, Los Angeles, CA 90095, USA}
\altaffiltext{7}{Lunar and Planetary Laboratory, University of Arizona, Tucson AZ 85721 USA}
\altaffiltext{8}{Subaru Telescope, NAOJ, 650 North A’ohoku Place, Hilo, HI 96720, USA}
\altaffiltext{9}{Department of Astrophysical Sciences, Princeton University, Princeton, NJ 08544, USA}
\altaffiltext{10}{Gemini Observatory, Casilla 603, La Serena, Chile}
\altaffiltext{11}{Large Binocular Telescope Observatory, 933 N Cherry Ave., Tucson AZ 85721, USA}
\altaffiltext{12}{Dunlap Institute for Astronomy \& Astrophysics, University of Toronto, 50 St. George St, Toronto ON M5S 3H4, Canada}
\altaffiltext{13}{Department of Physics and Astronomy, University of Georgia, Athens, GA 30602, USA}
\altaffiltext{14}{Institut de Recherche sur les Exoplan\`{e}tes, D\'{e}partment de Physique, Universit\'{e} de Montr\'{e}al, Montr\'{e}al QC H3C 3J7, Canada}
\altaffiltext{15}{Universit\'e Grenoble Alpes / CNRS, Institut de Plan\'etologie et d'Astrophysique de Grenoble, 38000 Grenoble, France}
\altaffiltext{16}{Physics \& Astronomy Department, Johns Hopkins University, 3600 N. Charles St., Baltimore MD, 21218, USA}
\altaffiltext{17}{European Southern Observatory, Alonso de Cordova 3107, Vitacura, Santiago, Chile}
\altaffiltext{18}{Large Synoptic Survey Telescope, 950 N Cherry Ave., Tucson AZ 85719, USA}
\altaffiltext{19}{Center for Astrophysics and Space Sciences, University of California, San Diego, La Jolla, CA 92093, USA}
\altaffiltext{20}{SETI Institute, Carl Sagan Center, 189 Bernardo Avenue, Mountain View, CA 94043, USA}
\altaffiltext{21}{NASA Ames Research Center, MS 245-3, Moffett Field, CA, 94035, USA}
\altaffiltext{22}{National Research Council of Canada Herzberg, 5071 West Saanich Road, Victoria, BC V9E 2E7, Canada}
\altaffiltext{23}{University of Victoria, 3800 Finnerty Rd, Victoria, BC, V8P 5C2, Canada}
\altaffiltext{24}{Department of Physics and Astronomy, Centre for Planetary Science and Exploration, The University of Western Ontario, London, ON, N6A 3K7, Canada}
\altaffiltext{25}{Stony Brook University, 100 Nicolls Rd, Stony Brook, NY 11794, USA}
\altaffiltext{26}{American Museum of Natural History, New York, NY 10024, USA}
\altaffiltext{27}{Infrared Processing and Analysis Center, California Institute of Technology, 700 S. Wilson Avenue, Pasadena, CA 91106, USA} 
\altaffiltext{28}{School of Earth and Space Exploration, Arizona State University, PO Box 871404, Tempe, AZ 85287, USA}
\altaffiltext{29}{Sibley School of Mechanical and Aerospace Engineering, Cornell University, Ithaca, NY 14853, USA}
\altaffiltext{30}{Jet Propulsion Laboratory, California Institute of Technology, 4800 Oak Grove Dr., Pasadena CA 91109, USA}
\altaffiltext{31}{The Aerospace Corporation, 2310 E. El Segundo Blvd., El Segundo, CA 90245}

\email{jwang@astro.berkeley.edu}

\begin{abstract}

A principal scientific goal of the Gemini Planet Imager (GPI) is obtaining milliarcsecond astrometry to constrain exoplanet orbits. However, astrometry of directly imaged exoplanets is subject to biases, systematic errors, and speckle noise. Here we describe an analytical procedure to forward model the signal of an exoplanet that accounts for both the observing strategy (angular and spectral differential imaging) and the data reduction method (Karhunen-Lo\`eve Image Projection algorithm). We use this forward model to measure the position of an exoplanet in a Bayesian framework employing Gaussian processes and Markov chain Monte Carlo (MCMC) to account for correlated noise. In the case of GPI data on $\beta$ Pic b, this technique, which we call Bayesian KLIP-FM Astrometry (BKA), outperforms previous techniques and yields 1$\sigma$-errors at or below the one milliarcsecond level. We validate BKA by fitting a Keplerian orbit to twelve GPI observations along with previous astrometry from other instruments. The statistical properties of the residuals confirm that BKA is accurate and correctly estimates astrometric errors. Our constraints on the orbit of $\beta$ Pic b firmly rule out the possibility of a transit of the planet at 10-$\sigma$ significance. However, we confirm that the Hill sphere of $\beta$ Pic b will transit, giving us a rare chance to probe the circumplanetary environment of a young, evolving exoplanet. We provide an ephemeris for photometric monitoring of the Hill sphere transit event, which will begin at the start of April in 2017 and finish at the end of January in 2018. 
\end{abstract}


\maketitle

\section{Introduction}
Astrometry is an essential tool for characterizing directly imaged exoplanets and their physical relationship to other elements of the planetary system 
in which they reside. The Gemini Planet Imager \citep[GPI;][]{macintosh14} was designed with a goal of achieving $\leq$1.8~mas astrometric accuracy \citep{graham09a}, necessary for characterizing the eccentricity distribution of exoplanet orbits from the GPI Exoplanet Survey \citep{konopacky14}. To do so, the astrometric calibration of GPI has continually been benchmarked to well-calibrated astrometric fields \citep{konopacky14}. This had led to some of the most precise astrometry on directly-imaged exoplanet systems to date \citep{mmb15, derosa15, rameau16}, allowing us to constrain or fit the first ever orbit of some of these directly-imaged exoplanets. However, limited by either the signal to noise ratio (SNR) of these exoplanets or by biases in the various data analysis algorithms, so far no astrometric study with GPI has achieved the design goal of 1.8~mas precision. 

The importance of understanding planetary orbits is highlighted by the $\beta$ Pictoris system, a young \citep[$\sim23$ Myr;][]{mamajek14, binks16} and nearby \citep[19.3~pc;][]{2007A&A...474..653V} system that has been extensively studied. $\beta$ Pic harbors a near edge-on debris disk that was first imaged by \citet{smith84} and which was subsequently observed to have a warp thought to be induced by a planet whose orbit is inclined relative to the debris disk \citep{burrows95a,mouillet97a,heap00a}. Additional indirect signatures of a planet were derived from variable spectral features modeled as infalling comets \citep{beust00a} and a peculiar light curve anomaly detected in 1981 \citep{lecavelier97}.

\citet{lagrange09a,lagrange10} then discovered \betapicb{}, a planet at an appropriate mass ($\sim10$~$M_{\text{Jup}}$) and semi-major axis ($8-13$~AU) to be responsible for the previously observed indirect signatures of planets.
A key focus of subsequent observations was determining the alignment of \betapicb{} relative to the main outer disk and the warp to determine if \betapicb{} is causing the warp. By observing the disk and planet simultaneously, \citet{lagrange12} concluded that the planet is misaligned from the main outer disk and consistent with being responsible for the warp. Additionally, \citet{dawson11} ruled out the possibility of having another giant planet in the system massive enough to cause the warp instead. Thus, \betapicb{} is responsible for the warp in the debris disk. This was confirmed in astrometric monitoring campaigns \citep{chauvin12, nielsen14, mmb15} which used homogeneous datasets to limit systematics and constrain the orbit of \betapicb{}.

Refining the orbital elements of $\beta$ Pic b
is not only crucial for investigating the dynamical link between the planet and the disk warp, but also because $\beta$ Pic b may transit its host star once every $\sim$20 years.  To date, there are no other known systems where the physical properties of an exoplanet can be characterized by using both the direct imaging and transit techniques.  Currently, the tightest constraint on the probability of transit is $\sim$0.06\%, obtained with a dedicated astrometric monitoring campaign with GPI \citep{mmb15}. However, \citet{lecavelier16} point out that GPI measurements from \citet{mmb15} have a higher position angle than astrometry from previous measurements, which could arise from a possible systematic calibration offset between GPI and previous instruments instead of actually due to \betapicb{}'s orbit being slightly inclined away from edge on. We note there currently is no evidence of a position angle offset with the GPI astrometric calibration, and the GPI astrometry of HD 95086 b are consistent with previous astrometry from other instruments \citep{rameau16}. 
However, it is important to more accurately compute the orbit of \betapicb{} because in late 2017 to early 2018 \citep{mmb15}, it will be at its closest projected separation from the star. The transit of the planet and/or any circumplanetary material orbiting around it could therefore be detectable.

One of the obstacles in characterizing directly imaged exoplanets is that even with the newest instrumentation, the glare of the host star covers the signal of the planet. In order to subtract the point spread function (PSF) of the star and maximize the SNR of the planet, observing techniques such as Angular Differential Imaging \cite[ADI;][]{marois06} and Spectral Differential Imaging \citep[SDI;][]{marois00} and data reduction algorithms like Karhunen-Lo\`eve Image Projection \citep[KLIP;][]{soummer12,pueyo15a} are used in combination to disentangle the PSF of the star from potential astrophysical sources. However, these techniques distort the planet signal and create data reduction artifacts, which usually are nuisance parameters that need to be calibrated out to obtain unbiased astrometry. 

Forward modelling effects of observing techniques and data reduction algorithms on the PSF of the planet was first done in the context of ADI and LOCI, showing significant improvements in astrometry and photometry for simulated planets \citep{marois10,galicher11}. In similar contexts with ADI and LOCI, \citet{brandt13} and \citet{esposito14} used forward modelling to correct for flux loss of exoplanets and the flux and morphology of disks respectively. For classical ADI (cADI), \citet{cantalloube15} showed that forward modelling techniques can improve the sensitivity of cADI and mitigate biases. The use of Markov-Chain Monte Carlo (MCMC) in conjunction with forward modelling was presented in \citet{bottom15} for reference differential imaging.

Recently, \citet{pueyo16} introduced a method, called KLIP-FM, to analytically forward model the degradation of a faint astrophysical signal that occurs when using least squares-based PSF subtraction algorithms such as KLIP that is also generally applicable to any observing strategy. Additionally, the computation of the forward model with KLIP-FM is much quicker than negative fake planet injection methods \citep{marois10, lagrange10}, as the stellar PSF subtraction algorithm, KLIP, needs only to be run once. In this paper, we demonstrate the advantages of KLIP-FM for precise astrometry and constraining planetary orbits by applying it to GPI observations of \betapicb{} reduced using KLIP and ADI+SDI. In Section \ref{sec:obs}, we describe our new astrometry technique, Bayesian KLIP-FM Astrometry (BKA),
in which we forward model the PSF of the planet with KLIP-FM and then use the forward model in a Bayesian framework to measure the position of the planet while also modelling the correlated nature of the noise. In Section \ref{sec:valid}, we validate our technique by fitting an orbit to the data and showing that our astrometry and uncertainties are consistent with Keplerian motion with no obvious systematics. Finally, in Section \ref{sec:orbit}, we apply our new astrometry to constrain the orbit of \betapicb{} and place the tightest constraints yet on the transit of the planet and its Hill sphere.

\section{Observations and Data Reduction}\label{sec:obs}
\subsection{Observations}
To obtain a large temporal baseline of GPI astrometric points, we compiled GPI data of $\beta$ Pic from commissioning (Gemini program GS-ENG-GPI-COM), an astrometric monitoring campaign of \betapicb{} (Gemini programs GS-2015A-Q-21 and GS-2015B-Q-9; PI: Graham), the GPI Exoplanet Survey (GS-2014B-Q-500; PI: Macintosh), and a Gemini Large and Long Program to characterize debris disks (GS-2015B-LP-6: PI: Chen). All the data used in the following analysis are listed in Table \ref{table:obs}. Most of these data were published and analyzed in a previous study characterizing the orbit of \betapicb{} by \citet{mmb15}. We have reprocessed those data with the new astrometry methods presented in this paper and combined them with five additional new epochs.

There were three datasets in \cite{mmb15} that we did not use. 
We did not use the polarimetry dataset as there was no instrumental PSF obtained with the data to forward model. The dataset on 2014 March 23 was taken during tests of the adaptive optics (AO) system, causing the instrumental PSF to be highly varying and making it unsuitable for forward modelling. The 2015-01-24 dataset contained artifacts in the construction of the forward model that we could not remove.
We chose not to include the measured astrometry from \citet{mmb15} for the datasets we omitted in order to maintain homogeneity in astrometric datasets and reduce potential systematic errors.
We also note that in the last dataset taken on 2016-01-21, there was saturation on the edge of the occulting mask due to bad seeing. This affected the forward modelling of \betapicb{}, which was also near the occulting mask so a large portion of the frames taken could not be used in this analysis.

\begin{deluxetable*}{clcccc|cccc}
\tabletypesize{\footnotesize}
\tablewidth{0pt} 
\tablecaption{Observations and Data Reduction Parameters for GPI Data on $\beta$ Pic\label{table:obs}} 
\tablehead{ 
 UT date & Program & \shortstack[c]{Filter} & \shortstack[c]{Exposure \\ time\\ (s)} & \shortstack[c]{Field \\ rotation \\ (\degr)} & \shortstack[c]{Average \\seeing\tablenotemark{a} \\ (\arcsec)} & \shortstack[c]{KL \\ modes} & \shortstack[c]{Exclusion \\ criteria \\ (pixels)} & \shortstack[c]{High-pass \\ filtered } & \shortstack[c]{Fitting \\ box  size \\ (pixels)}
}
\startdata 
2013 Nov 16 & GS-ENG-GPI-COM & \textit{K1} & 1789 & 26 & 1.09 & 7 & 4 & No  & 13  \\
2013 Nov 16 & GS-ENG-GPI-COM & \textit{K2} & 1253 & 18 & 0.93  & 7 & 4 & No & 13\\
2013 Nov 18 & GS-ENG-GPI-COM & \textit{H}  & 2446 & 32 & 0.68   & 7 & 4 & No & 13\\
2013 Dec 10 & GS-ENG-GPI-COM & \textit{H}  & 1312 & 39 & 0.77 & 7 & 4 & No & 13 \\
2013 Dec 10 & GS-ENG-GPI-COM & \textit{J}  & 1597 & 19 & 0.70 & 7 & 4 & No & 13 \\
2013 Dec 11 & GS-ENG-GPI-COM & \textit{H}  & 410  & 65 & 0.46 & 7 & 4 & No & 11 \\
2014 Nov 8 & GS-ENG-GPI-COM & \textit{H} & 2147 & 25 & 0.77 & 7 & 4 & No & 11 \\
2015 Apr 2 & GS-2015A-Q-21 & \textit{H} & 1312 & 10 & 0.51 & 2 & 2 & No & 11 \\
2015 Nov 6 & GS-2014B-Q-500 & \textit{H} & 2207 & 28 & -\tablenotemark{b} & 7 & 4 & No  & 11 \\
2015 Dec 5 & GS-2015B-LP-6 & \textit{J} & 4948 & 66 & 0.92  & 7 & 4 & Yes  & 11\\ 
2015 Dec 22 & GS-2015B-Q-9 & \textit{H} & 2088 & 19 & 0.76 & 7 & 4 & No  & 11 \\
2016 Jan 21 & GS-2015B-Q-9 & \textit{H} & 954 & 17 & 1.18 & 10 & 2 & Yes & 7  \\
\enddata
\tablenotetext{a}{Measured by the Gemini DIMM}
\tablenotetext{b}{Seeing monitor data were not available for this observation} 
\end{deluxetable*}

\subsection{Reducing Raw GPI Data}
The raw integral field spectrograph (IFS) data were reduced to form 3-D spectral data cubes using the GPI Data Reduction Pipeline (DRP) version 1.2.1 or 1.3 \citep{perrin14_drp}. There were no significant changes between the two versions of the GPI DRP that impacted the astrometry of \betapicb{}.
We used the same data reduction process as in \citet{mmb15} and will summarize them here.
First, dark subtraction and bad pixel correction were applied to each 2-D image. For earlier datasets in which cryocooler vibration induced correlated noise on the detector, the frames were ``destriped'' to remove this noise \citep{Ingraham14}. Then, we corrected for instrument flexure using an argon arc lamp taken before each sequence to align each individual spectrum for extraction \citep{wolff14}. Then each 2-D frame was turned into a spectral data cube, corrected for any remaining bad pixels, and corrected for distortion \citep{konopacky14}. For the \textit{K}-band data, thermal background frames were taken along the sequence. We constructed thermal background cubes in the same fashion and subtracted them from the \textit{K}-band data. 

To spatially register and photometrically calibrate our data, we used the GPI DRP to measure the flux and location of the ``satellite'' spots, which are centered on the occulted star and imprinted with its attenuated spectrum \citep{marois06b, sivaramakrishnan06,wang14}. The satellite spot fluxes were used to derive a flux calibration in each spectral channel, which we used when constructing the forward models in Section \ref{sec:fm}. 

The location of the occulted star at each wavelength in each datacube was found using a least squares fit to all of the satellite spots' positions and the magnitude of the atmospheric differential refraction. The occulted star center is used to align all the images together before PSF subtraction, and is crucial for determining the astrometry of \betapicb{} relative to its host star. The precision on the star center is 0.05 pixels (0.7~mas) for satellite spots with signal-to-noise ratios (SNR) $>$ 20 \citep[][]{wang14}, which is certainly the case for all of our data on bright stars like $\beta$ Pic.

\subsection{PSF Subtraction}\label{sec:psfsub}
To subtract the stellar PSF from each individual datacube, we used \texttt{pyKLIP} \citep{wang15}, a Python implementation of the KLIP algorithm. We used both ADI and SDI to decorrelate the stellar PSF from the PSF of \betapicb{}. As we were only concerned with \betapicb{}, we applied our PSF subtraction only on the annulus that included the planet, rather than the full image. 

We adjusted three main parameters for the PSF subtraction, depending on the dataset. The first was the number of modes used from the Karhunen-Lo\`eve (KL) transform to model the stellar PSF. The second was an exclusion criteria for picking reference PSFs. The exclusion criteria is similar to the quantity $N_{\delta}$ in \citet{laf07} and is defined by the number of pixels that \betapicb{} would move azimuthally and radially in an observing sequence due to ADI and SDI. Thirdly, we toggled an $\sim$11~pixel wide spatial high-pass filter that was applied to some datasets before PSF subtraction. The high-pass filter was implemented using a Gaussian filter in Fourier space and the $\sim$11~pixel size was chosen to remove low frequency background without significantly distorting any point sources in the image. 
In general, increasing the number of KL modes, decreasing the exclusion criteria, and applying a high-pass filter improves the subtraction of the stellar PSF. However, taken to the extreme, all three options attenuate signal from the planet. Additionally, for a planet as bright as \betapicb{}, the forward modelling described in Section \ref{sec:fm} may not be valid when the PSF subtraction becomes too aggressive \citep{pueyo16}. Thus, to optimize the signal of the planet while maintaining the validity of the forward modelling, we varied these parameters for each dataset. As \betapicb{} is bright, in most of the datasets we used an exclusion criteria of 4 pixels, which is slightly greater than 1~$\lambda/D$, where $\lambda$ is the wavelength and $D$ is the diameter of the telescope. When there was little field rotation for ADI, we decreased the exclusion criteria to not overly restrict our reference PSFs, but also decreased the number of KL modes used to avoid being too aggressive. When the speckle noise was too bright due to observing at a shorter wavelength or when the planet moved closer to the star, we applied a high pass filter to remove some of the diffracted starlight. We list the chosen parameters for each dataset in Table \ref{table:obs}.

After PSF subtraction, all images were rotated so that north is up and east is left. Then, all the frames were mean combined in both the spectral and temporal directions, resulting in a single PSF-subtracted frame for each dataset.

\subsection{Constructing the Forward Model}\label{sec:fm}
After stellar PSF subtraction, the PSF of the planet is distorted by over-subtraction, caused by the presence of the planet in the data we are subtracting, and self-subtraction, caused by the presence of the planet in the reference images. Over and self-subtraction perturbs the astrometric and photometric properties of the planet's PSF and prevents a straightforward measurement of its position. Typically, the biases and uncertainties in the astrometry caused by stellar PSF subtraction
are estimated by injecting fake planets into the data at other position angles and comparing the retrieved position to the injected position. 
However, the over-subtraction and self-subtraction that distorts the planet PSF is deterministically caused by the existence of a planet in the data and its apparent motion in the reference images induced by ADI and SDI. In turn, these features, if modeled, can inform us about the location of the planet and improve our astrometric precision and accuracy.

Recently, \citet{pueyo16} introduced KLIP-FM, an analytic framework to compute the effect of a planet on the KL modes and use these perturbations to reconstruct the over-subtraction and self-subtraction features. Using this technique, we are able to generate the PSF of the planet after PSF subtraction (see Appendix \ref{appendix:fmpsf} for a detailed description of the procedure). Briefly, we use the PSF of the planet before PSF subtraction, the apparent motion of the planet due to ADI and SDI, and a model of the planet spectrum to compute the distorted PSF of the planet after PSF subtraction. The result is a 2-D broadband planet PSF, which we call $F$, centered at ($x_0$, $y_0$), our initial estimate for the location of the planet.

For our GPI data, the forward models were generated using the implementation of KLIP-FM in \texttt{pyKLIP}.
For each dataset, we used the same parameters as our PSF subtraction to construct the forward model. 
To construct a model of the instrumental PSF at each wavelength, we used the average PSF of the satellite spots across all images for that wavelength. 
We used a basic center of light centroiding routine to measuring an approximate $x$ and $y$ positions of the planet ($x_0$ and $y_0$) in the PSF subtracted image. This initial estimate for the position is good to within a pixel for our GPI data. For the input planet spectrum of the forward model, we used the normalized best-fit model spectrum from \citet{chilcote15}, which has an effective temperature of 1650~K and a $\log(g)=4.0$ (cgs units).
We scaled the spectrum by eye to approximately match the contrast of \betapicb{} in our data and used the satellite spot fluxes to convert from contrast to digital numbers. We found that varying the spectrum and photometry had negligible impacts on our measured astrometry. Even a spectrum differing by $\sim$25\% in shape changed the astrometry by $< 0.2$~mas, significantly smaller than the uncertainties we find in the following analysis.

\begin{figure}[Ht]
\plotone{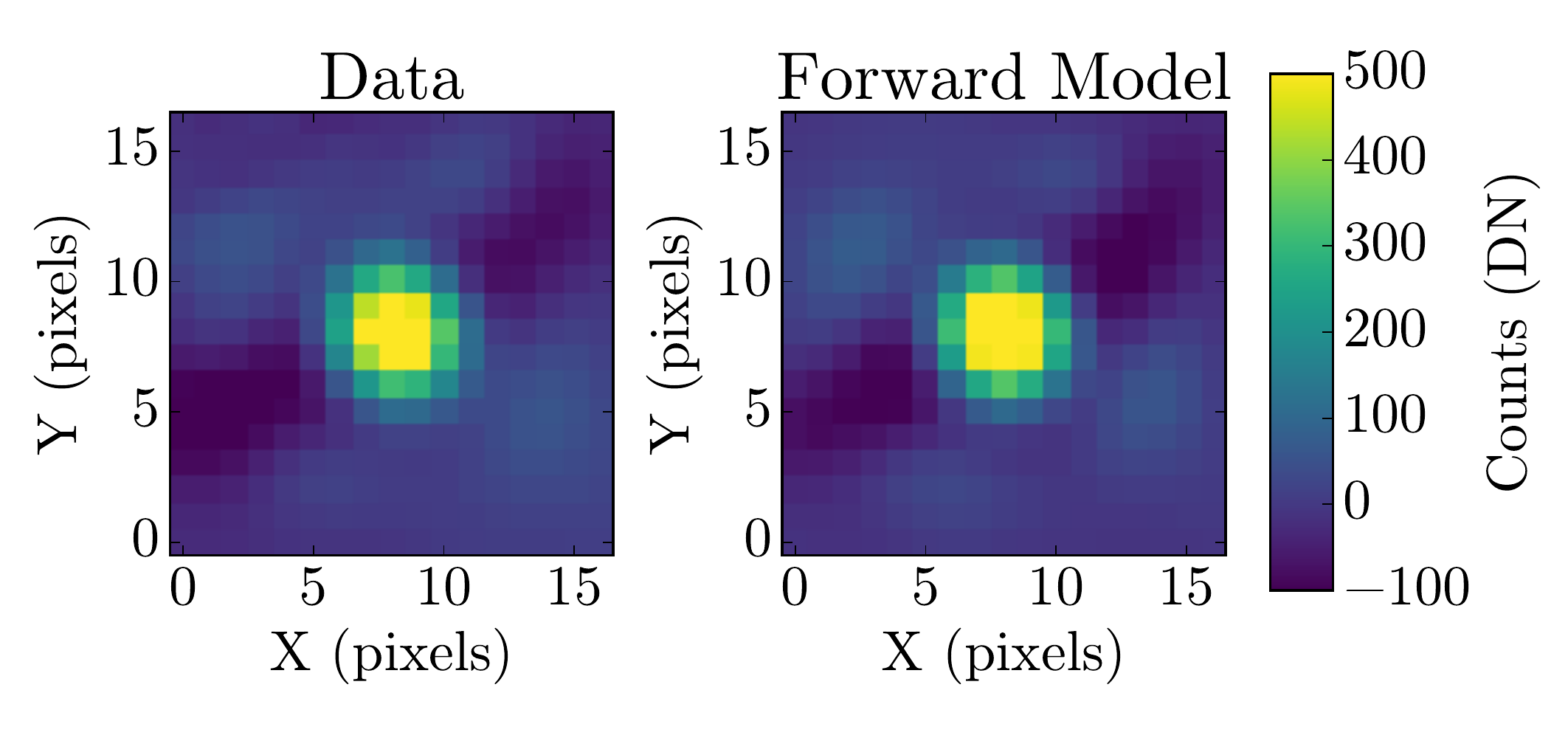}
\caption{\textit{(Left)} An image of \betapicb{} from the 2013 November 18 GPI \textit{H}-band data after stellar PSF subtraction. \textit{(Right)} Unoptimized forward model of \betapicb{} for the same dataset. The forward model has not been optimized to fit the data yet using the MCMC procedure discussed in Section \ref{section:mcmcastrom}, but should already be accurate to within a pixel. 
To see the best fit forward models and residuals for each GPI dataset, see Appendix \ref{appendix:res}. The star is to the upper left of the planet and far outside the region shown here. 
\label{fig:data-fm-compare}}
\end{figure}

With this input information, we used \texttt{pyKLIP} to generate one forward modeled PSF, $F$, for each dataset. In Figure \ref{fig:data-fm-compare}, we show an example forward modeled PSF (not optmized to fit the data) and comparison to data for our 2014 November 18 dataset on \betapicb{}.
Qualitatively, the forward model matches all the features seen in data, including the prominent negative self-subtraction lobes on either side of the planet. As \texttt{pyKLIP} parallelizes the computation, the generation of the forward models is quick. On a 32-core machine with AMD Opteron 6378 processors clocked at 2.3~GHz, forward models for all 37 channels of a representative 37-cube GPI dataset were generated in 15 minutes: 4 minutes of overheads for preprocessing and generating the instrumental PSFs and 11 minutes to execute KLIP-FM and create the forward model. We note that we chose to run the forward model on a large annulus to examine noise properties in the data and that the computation time for KLIP-FM decreases $\sim$20-30\% if a small sector around the planet was used instead.

\subsection{Locating the Planet with Bayesian Parameter Estimation}\label{section:mcmcastrom}
To use the forward modeled PSF, $F$, to perform astrometry, we developed a Bayesian framework to fit $F$ to the data, account for correlated noise, and estimate our fitting uncertainties. First, ignoring the correlated noise, we can use three parameters to fit the forward model to the data: the location on the planet in $x$ ($x_p$), the location of the planet in $y$ ($y_p$), and a scale factor ($\alpha$) to match the flux of $F$ to the data.
We can then write the posterior probability for $x_p$, $y_p$, and $\alpha$ given the data $D$ using Bayes' Theorem as
\begin{equation}
P(x_p,y_p,\alpha|D)  = P(D|x_p,y_p,\alpha)P(x_p,y_p,\alpha),
\end{equation}
ignoring normalization constants. The first term on the right-hand side is the likelihood and the second is the prior.

To construct the likelihood, we must first use our input parameters and $F$ to generate a model to compare to the data. 
We scale $F$ by $\alpha$ and recenter it from its guessed location ($x_0$, $y_0$) to ($x_p$, $y_p$). The residual, $R$, between the model and the data, $D$, in fitting region $\mathcal{F}$ is calculated by
\begin{equation}
R \equiv R(x_p, y_p, \alpha) = \left(D - \alpha F(x_p,y_p)  \right)_{\mathcal{F}},
\end{equation}
where fitting region $\mathcal{F}$ is a fixed rectangular box centered at the approximate location of the planet in the data. 
We pick the size of the fitting region to be a few $\lambda/D$ to encompass the PSF and the self-subtraction wings. We varied the fitting box size for each of our datasets in order to keep the fit focused on the area where signal from the planet can be seen.
We list the size of the fitting box for each dataset in Table \ref{table:obs}. Note that we do not fit any background term to the image, as one of the first steps of KLIP is to subtract off the mean of the image, removing any spatially constant background.

For data with uncorrelated errors, the log of the likelihood function of the data for a particular model is
\begin{equation}
\begin{aligned}
P(D|x_p,y_p,\alpha) &= \mathcal{\ln L}(x_p,y_p,\alpha) \\
&=  -\frac{1}{2} \sum_i^{N_{\text{pix}}} \left[ \frac{R_i^2(x_p,y_p,\alpha)}{\sigma_i^2} + \ln(2 \pi \sigma_i) \right],
\end{aligned}
\end{equation}
where $\sigma_i$ is the uncertainty in pixel i and $N_{\text{pix}}$ is the number of pixels in the fitting region. However, the assumption of uncorrelated errors does not hold for images limited by speckle noise. 
Except in the cases of very aggressive PSF subtraction or at much greater separations from the star, the residual noise after PSF subtraction in GPI data of a bright star like $\beta$ Pic is still dominated by correlated speckle noise, which has a correlation scale that depends on the aggressiveness of the PSF subtraction. Due to the bright nature of $\beta$ Pic b, we could not use a very aggressive reduction to ensure that the planet remained a perturbation on our KL modes, 
thereby preserving the validity of the analytical forward modelling technique. The conservative PSF subtraction combined with the close separation of \betapicb{} from its bright host star required us to capture the correlated nature of the noise.  
We thus write our likelihood function instead as
\begin{equation}
\mathcal{\ln L} =  -\frac{1}{2} (R^T C^{-1} R + \ln(\det C) + N_{\text{pix}} \ln(2\pi) ),
\end{equation}
where $C$ is the covariance matrix of size $N_{\text{pix}} \times N_{\text{pix}}$ and $R$ is a $N_{\text{pix}} \times 1$ matrix.

We applied a Gaussian process framework to characterize the covariance in the noise (see \citet{czekala15} for an in-depth explanation of the application of Gaussian processes to astronomical data). We only aim to model the correlations within each individual speckle, which spans $\lambda/D$ in spatial extent. While in a single unprocessed frame speckle noise also has addition correlations on much larger scales, there are no significant correlations between speckles at larger spatial scales in our small fitting region due to stellar PSF subtraction with KLIP and averaging uncorrelated speckles together when collapsing the frames in our ADI+SDI sequence. Thus, within our fitting region, the dominant correlation in our noise is from pixels within a single speckle, and thus is the one correlation we modelled.

Following the procedure in \citet{czekala15} for fitting 1-D correlations in stellar spectra, we used the Mat\'{e}rn covariance function parametrized with $\nu=3/2$ to model the correlated speckle noise. We chose the Mat\'{e}rn function with $\nu=3/2$ as it better fits the correlations at larger separations compared to a simple squared exponential relation. However, we note that switching between covariance functions that have similar shapes does not significantly alter the error bars. We also chose to assume symmetric noise as we did not find any difference in the correlation scale of our noise in the radial and azimuthal directions. This is likely be due to the fact that we used both ADI and SDI to model the stellar PSF and thus are better at subtracting speckle noise, which before PSF subtraction is more correlated radially than azimuthally due to the finite spectral bandwidth of the data.
In instruments that are not able to utilize SDI, it might be necessary to separate the noise into radial and azimuthal components, each with its own characteristic correlation length. Thus for our purposes, we chose the symmetric Mat\'{e}rn covariance function with $\nu=3/2$, which defines the covariance between two pixels $i$ and $j$ separated by a distance of $r_{ij} = \sqrt{(x_i-x_j)^2 + (y_i-y_j)^2}$ as
\begin{equation}
C_{ij} = \sigma_i \sigma_j \left( 1+ \frac{\sqrt{3} r_{ij}}{l} \right) \exp\left( -\frac{\sqrt{3} r_{ij}}{l} \right),
\end{equation}
where $\sigma_i$ and $\sigma_j$ are the uncertainties for each pixel and $l$ is a characteristic correlation length scale that increases when noise is correlated at larger spatial scales. 
We calculated the uncertainty for each pixel by computing the standard deviation of pixels in an annulus with a width of $\sim2\lambda/D$ (6~pixels), centered on the star, and with a mean radius equal to the distance between that pixel and the star. Any pixels containing signal from the planet were masked and not used in estimating the noise in the annulus.

\begin{figure*}[!ht]
\epsscale{.75}
\plotone{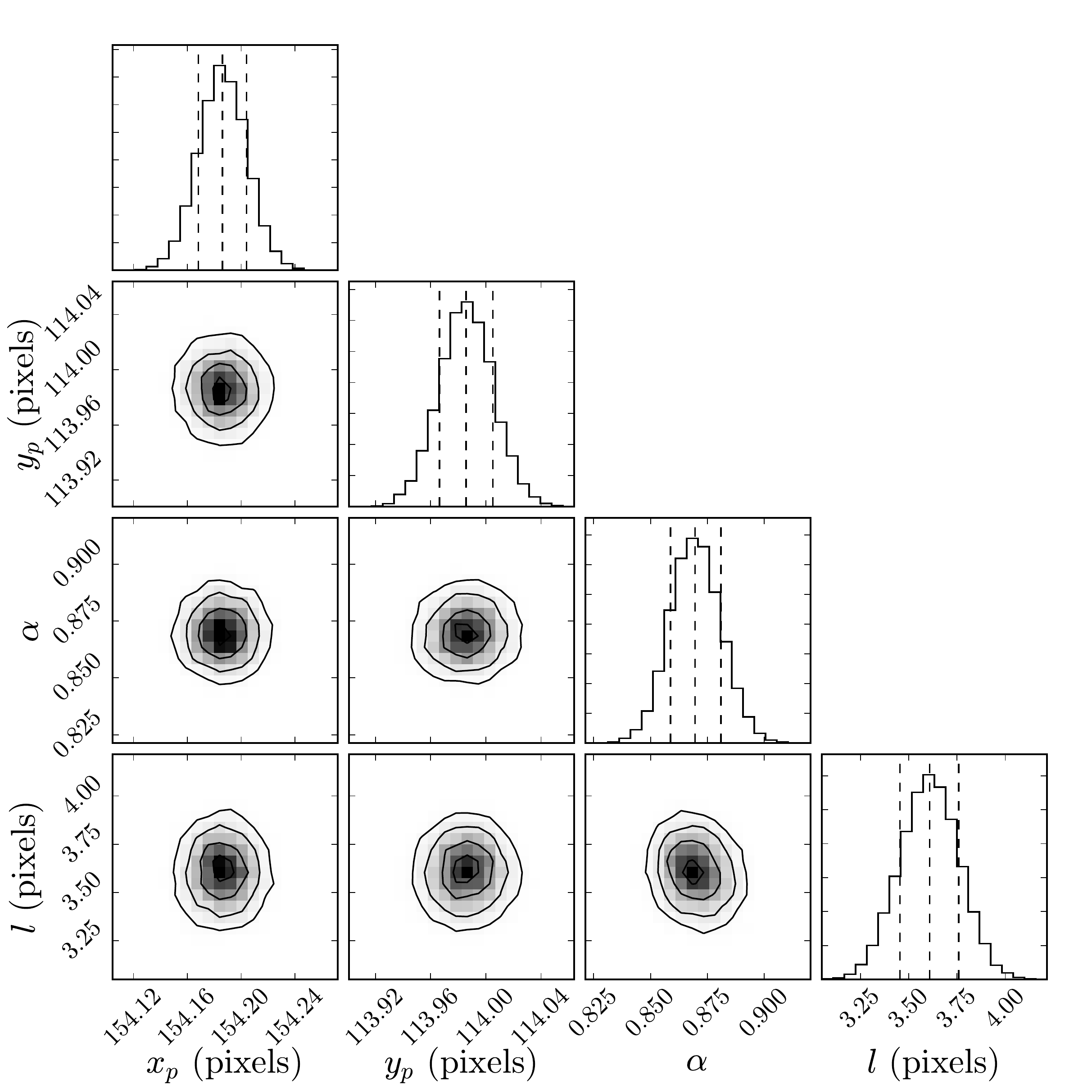}
\caption{ Posterior distribution of the four parameters in the MCMC fit for the astrometry for the 2014 November 18 epoch. The vertical dashed lines in the marginalized posterior distribution plots indicate the 16th, 50th, and 84th percentile values.
\label{fig:astrom-tri}}
\end{figure*}

The correlation length $l$ is not fixed, but rather kept as a hyperparameter parameter in our Bayesian framework that we will marginalize over in the end. Thus the final posterior we are trying to calculate is really
\begin{equation}
P(x_p,y_p,\alpha,l|D)  = P(D|x_p,y_p,\alpha,l)P(x_p,y_p,\alpha,l),
\end{equation}
where $\alpha$ and $l$ are both hyperparameters as we are only interested in the astrometry.

Compared to our likelihood function, our prior is relatively straightforward. We allow for uniform priors in $x$ and $y$ within 2 pixels from our initial guess location. Similarly, we allow for an uniform prior between 0 and 5 for $\alpha$ to determine how much to scale $F$, which was already scaled to an approximate contrast of \betapicb{}. 
The correlation length $l$ has a uniform prior between 0 and 10 pixels, which provides an ample range to explore the correlations within individual speckles of size $\lambda/D\approx 3$~pixels.

We used the \citet{goodman10} Affine Invariant MCMC sampler implemented in the \texttt{emcee} Python package \citep{emcee13} to sample the posterior distribution and custom \texttt{cython} code to quickly generate the covariance matrix as we vary $l$. The MCMC sampler was run for 800 steps using 100 walkers, with a "burn-in" of 200 steps beforehand. In Figure \ref{fig:astrom-tri}, we show the posterior distributions from the MCMC fit for the 2014 November 18 \textit{H}-band dataset as a representative posterior distribution. The value of $l$ is close to $\lambda/D \approx 3$~pixels, indicating we are accurately fitting the correlated speckle noise. We use the resulting posterior distributions to calculate the most-likely values and uncertainties for the location of \betapicb{} in our image at each epoch.

To convert our results to more useful physical values, we need to convert our measured location of \betapicb{} to its positional offset from its host star in right ascension (RA) and declination (Dec). As our images are already rotated so that $-x$ is positive RA and $+y$ is positive Dec, it is straightforward to convert from image fates to sky coordinates. We use the satellite spots to measure the location of the star behind the coronagraph, which has a precision of 0.7~mas \citep{wang14}. 
This allows us to derive the separation of the planet from the star in pixels. To convert from pixel separations to physical separations in RA and Dec, we use the most-recent astrometric calibration numbers from \citet{derosa15}: a plate scale of $14.166 \pm 0.007$ mas~lenslet$^{-1}$ and a residual North angle offset of $0\fdg10 \pm 0\fdg13$ from the North angle value used in the GPI DRP (versions 1.2.1 to 1.3, the current version). These astrometric calibration numbers show no significant change over time, so we apply them to all our epochs of data. Then, we assume all of these error terms are uncorrelated and add them in quadrature with our measurement errors from our MCMC fit to determine our full astrometric precision. 

The combination of the forward modeled PSF and the Bayesian framework makes up the Bayesian KLIP-FM Astrometry technique we introduce in this work. We apply BKA for all twelve GPI datasets and report the measured astrometry and error budgets in Table \ref{table:astrometry}. The best fit forward models and residuals to the fit are shown in Appendix \ref{appendix:res}. 
On most of our datasets, we are not limited just by the uncertainty in the location of the planet, which was as low as 0.3~mas. The uncertainties in the location of the star and North angle also make significant contributions to the error budget. Typically, we achieved $\sim$1~mas precision on the relative astrometry between \betapicb{} and its host star. This is a factor of $\sim$2--4 improvement over previous techniques \citep{mmb15} using the same data, indicating this technique can be useful in reanalysis of archival data to obtain better astrometry in cases where the limiting factor is the uncertainty on the planet position. In two of the later datasets where the planet is observed closer in, we were limited by the SNR of the planet and unable to achieve 1~mas precision. In the 2015 December 5 dataset, the noise was higher due to the planet being fainter relative to the star in \textit{J}-band. In the 2016 January 21 dataset, a combination of poor seeing and a small amount of usable data limited our astrometric precision. 

Overall though, this GPI \betapicb{} data is an excellent demonstration for Bayesian KLIP-FM Astrometry as the planet is bright enough that the extended PSF features, such as the negative self-subtraction lobes, are clearly seen and provide significant information to constrain the position of the planet. For fainter planets, the extended features are harder to distinguish from the noise. As one of the main advantages of BKA over techniques that do not forward model the PSF is being able to forward model the extended self-subtraction lobes, the astrometric improvement would not be as large for lower signal-to-noise ratio planets.
There still should be some improvement though due to accurately modelling the over-subtraction on the core of the PSF and small contributions from the extended features even if they are hard to distinguish from noise. Regardless, in addition to the improved precision, BKA should also more accurately estimate the uncertainties as it fits for the correlation scale of the noise at the location of the planet. 

\begin{deluxetable*}{lcccc|cccc}
\tablecaption{Astrometric Error Budget 
and Measured Astrometry of $\beta$ Pic b \label{table:astrometry} }
\tablewidth{0pt} 
\tablehead{ 
Dataset  &  \shortstack[c]{Planet $x$/$y$ \\ Uncertainty \\ (mas)}  &  \shortstack[c]{Star $x$/$y$ \\ Uncertainty \\ (mas)}  &  \shortstack[c]{Plate Scale \\ Uncertainty \\ (mas)}  &  \shortstack[c]{PA \\ Uncertainty \\ (\degr)} &  \shortstack[c]{$\Delta$RA \\ (mas)} & \shortstack[c]{$\Delta$Dec \\ (mas)} & \shortstack[c]{Radial \\ Separation (mas) } & \shortstack[c]{Position Angle \\  (\degr)}
}
 \startdata
2013 Nov 16 $K1$ & 0.6/0.7 & 0.7/0.7 & 0.3 & 0.13 & -228.5 $\pm$ 1.3 & -366.2 $\pm$ 1.1 & 431.6 $\pm$ 1.0 & 212.0 $\pm$ 0.2 \\
2013 Nov 16 $K2$ & 0.5/0.4 & 0.7/0.7 & 0.3 & 0.13 & -229.2 $\pm$ 1.2 & -364.5 $\pm$ 1.0 & 430.6 $\pm$ 0.9 & 212.2 $\pm$ 0.2 \\
2013 Nov 18 $H$ & 0.3/0.3 & 0.7/0.7 & 0.3 & 0.13 & -229.1 $\pm$ 1.1 & -364.7 $\pm$ 1.0 & 430.6 $\pm$ 0.8 & 212.1 $\pm$ 0.2 \\
2013 Dec 10 $H$ & 0.4/0.4 & 0.7/0.7 & 0.3 & 0.13 & -227.9 $\pm$ 1.2 & -359.9 $\pm$ 1.0 & 426.0 $\pm$ 0.9 & 212.3 $\pm$ 0.2 \\
2013 Dec 10 $J$ & 0.6/0.7 & 0.7/0.7 & 0.3 & 0.13 & -227.2 $\pm$ 1.3 & -361.1 $\pm$ 1.2 & 426.6 $\pm$ 1.1 & 212.2 $\pm$ 0.2 \\
2013 Dec 11 $H$ & 0.5/0.4 & 0.7/0.7 & 0.3 & 0.13 & -227.8 $\pm$ 1.2 & -359.2 $\pm$ 1.0 & 425.4 $\pm$ 0.9 & 212.4 $\pm$ 0.2 \\
2014 Nov 8 $H$ & 0.5/0.5 & 0.7/0.7 & 0.2 & 0.13 & -194.0 $\pm$ 1.1 & -299.1 $\pm$ 1.0 & 356.5 $\pm$ 0.9 & 213.0 $\pm$ 0.2 \\
2015 Apr 2 $H$ & 0.4/0.5 & 0.7/0.7 & 0.2 & 0.13 & -172.1 $\pm$ 1.0 & -266.5 $\pm$ 1.0 & 317.2 $\pm$ 0.9 & 212.9 $\pm$ 0.2 \\
2015 Nov 6 $H$ & 0.7/0.7 & 0.7/0.7 & 0.2 & 0.13 & -137.8 $\pm$ 1.1 & -207.2 $\pm$ 1.0 & 248.8 $\pm$ 1.0 & 213.6 $\pm$ 0.3 \\
2015 Dec 5 $J$ & 1.2/1.3 & 0.7/0.7 & 0.2 & 0.13 & -133.9 $\pm$ 1.5 & -200.5 $\pm$ 1.5 & 241.1 $\pm$ 1.4 & 213.7 $\pm$ 0.4 \\
2015 Dec 22 $H$ & 0.5/0.5 & 0.7/0.7 & 0.2 & 0.13 & -130.0 $\pm$ 1.0 & -194.7 $\pm$ 0.9 & 234.1 $\pm$ 0.9 & 213.7 $\pm$ 0.2 \\
2016 Jan 21 $H$ & 1.8/1.6 & 0.7/0.7 & 0.2 & 0.13 & -126.8 $\pm$ 2.0 & -185.8 $\pm$ 1.8 & 225.0 $\pm$ 1.8 & 214.3 $\pm$ 0.5 \\
\enddata
\end{deluxetable*}

\begin{deluxetable*}{ccccccc}
\tablecaption{Orbital Parameters of $\beta$ Pic b \label{tab:orb_params}} 
\tablehead{ 
&&&& \multicolumn{3}{c}{Posterior Percentiles} \\ \cline{5-7} \\
Parameter & Unit & Prior Range & Prior Distribution & 16 & 50 & 84}
\startdata
Semi-major axis ($a$) & au & $4 - 40$ & Uniform in $\log a$ & 9.02 & 9.66 & 10.78 \\
Epoch of Periastron ($\tau$) & - & $-1.0 - 1.0$ & Uniform in $\tau$ & 0.32 & 0.73 & 0.87 \\
Argument of Periastron ($\omega$) & \degr & $-360 - 360$ & Uniform in $\omega$ & 192.8 & 205.8 & 258.4 \\
Position Angle of the Ascending Node ($\Omega$) & \degr & $25 - 85$  & Uniform in $\Omega$ & 31.67 & 31.76 & 31.84\\
Inclination ($i$) & \degr & $ 81 -99$ & Uniform in $\cos{i}$ & 88.70 & 88.81 & 88.93 \\
Eccentricity ($e$) & - & $0.00001 - 0.99$ & Uniform in $e$ & 0.027 & 0.080 & 0.171 \\
Total Mass ($M_T$) & $M_{\odot}$ & $0 - 3$ & Uniform in $M_{\odot}$ & 1.76 & 1.80 & 1.83 \\
\hline
\multicolumn{7}{c}{Derived Parameters} \\
\hline
Period ($P$) & years & - & - & 20.21 & 22.47 & 26.24 \\
Hill Sphere Ingress & MJD & - & - & 57,840 & 57,846 & 57,854 \\
$1/2$ Hill Sphere Ingress & MJD & - & - & 57,916 & 57,924 & 57,934 \\
Closest Approach Date & MJD & - & - & 57,986 & 57,996 & 58,008 \\
$1/2$ Hill Sphere Egress & MJD & - & - & 58,056 & 58,069 & 58,082 \\
Hill Sphere Egress & MJD & - & - & 58,132 & 58,147 & 58,162 \\
\enddata
\end{deluxetable*}

\begin{figure*}[!ht]
\plotone{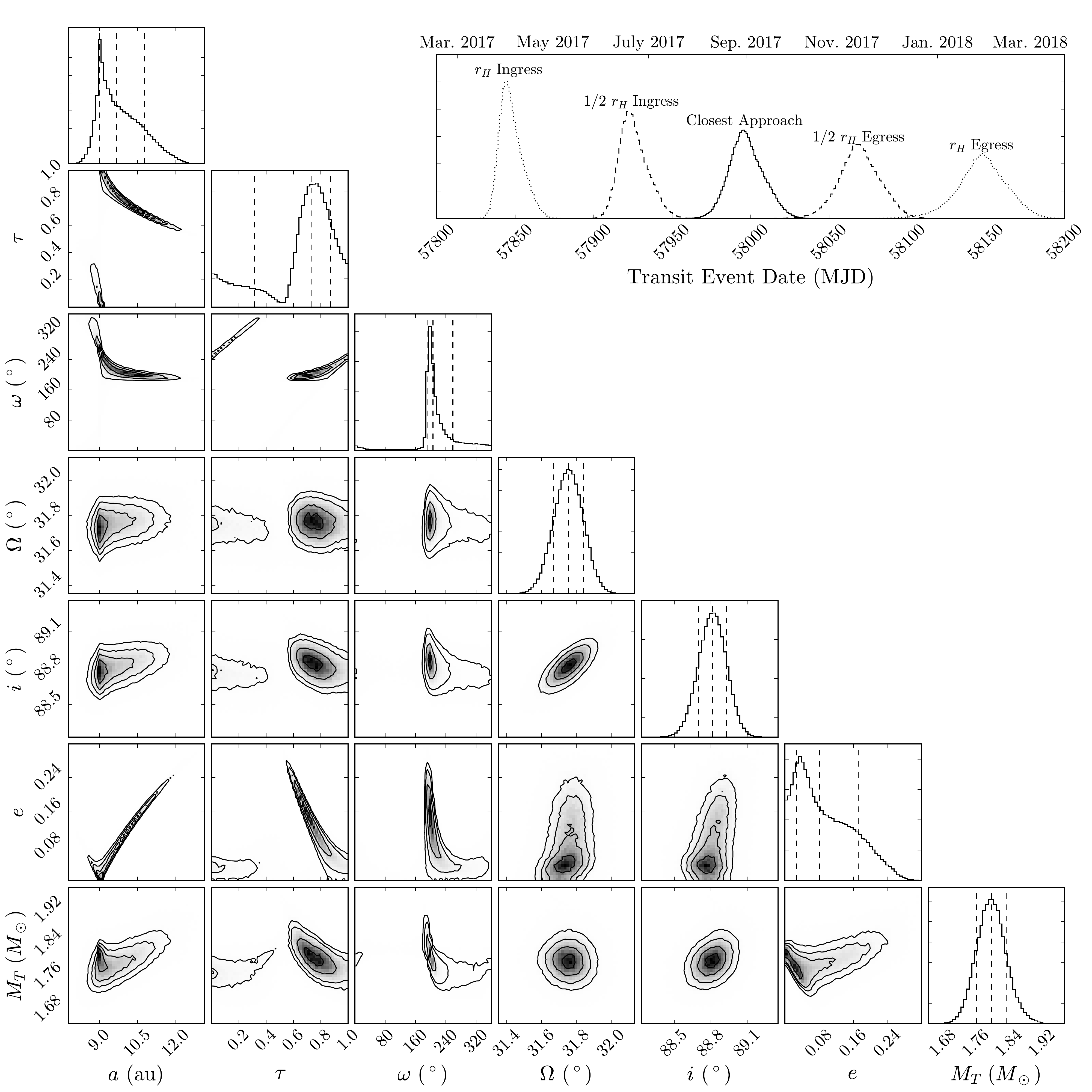}
\caption{Posterior distributions for the seven orbital elements in our Keplerian orbit model along with inferred distributions of possible dates for transit events in the top right corner. The vertical dashed lines in the marginalized posterior distribution plots indicate the 16th, 50th, and 84th percentile values. For the transit events, the dotted line corresponds to the ingress and egress of the full Hill sphere, the dashed line corresponds to the ingress and egress of the half Hill sphere, and the solid line corresponds to the date of closest approach. \label{fig:orbit_triangle}}
\end{figure*}

\section{Validation through Orbit Fitting}\label{sec:valid}
To explore the validity of our new astrometry technique, we fit a Keplerian orbit to our twelve epochs of astrometry. Since each epoch of astrometry is fit independently, and since the Keplerian orbit is agnostic towards the exact data reduction methods, having all twelve epochs of astrometry fit the Keperlain orbit would only be possible if all of the astrometry is accurate and precise. If there are errors in estimating the magnitude of the uncertainties or any remaining biases in our measurements, we expect them to become evident in the residuals of the Keplerian fit either as systematic trends or fit outliers.

We followed the same analysis as in \citet{mmb15} to obtain the orbital elements of \betapicb{} using a MCMC fit with the parallel-tempered sampler implemented in \texttt{emcee}. We combined the twelve GPI astrometric points presented here along with the datasets presented in \citet{chauvin12} and \citet{nielsen14}. Unlike \citet{mmb15}, we did not explicitly include the radial velocity measurement of \betapicb{} from \citet{snellen14} in order to limit potential systematics between instruments, but we do use it to constrain the prior on the position angle of the ascending node and thus the direction of the orbit (i.e., we know that that \betapicb{} has been moving towards us since $\sim$2007). Our model of the orbit fits the same seven parameters as \citet{mmb15}. For convenience, the parameters are listed in Table \ref{tab:orb_params}. 

\begin{figure*}[!ht]
\plotone{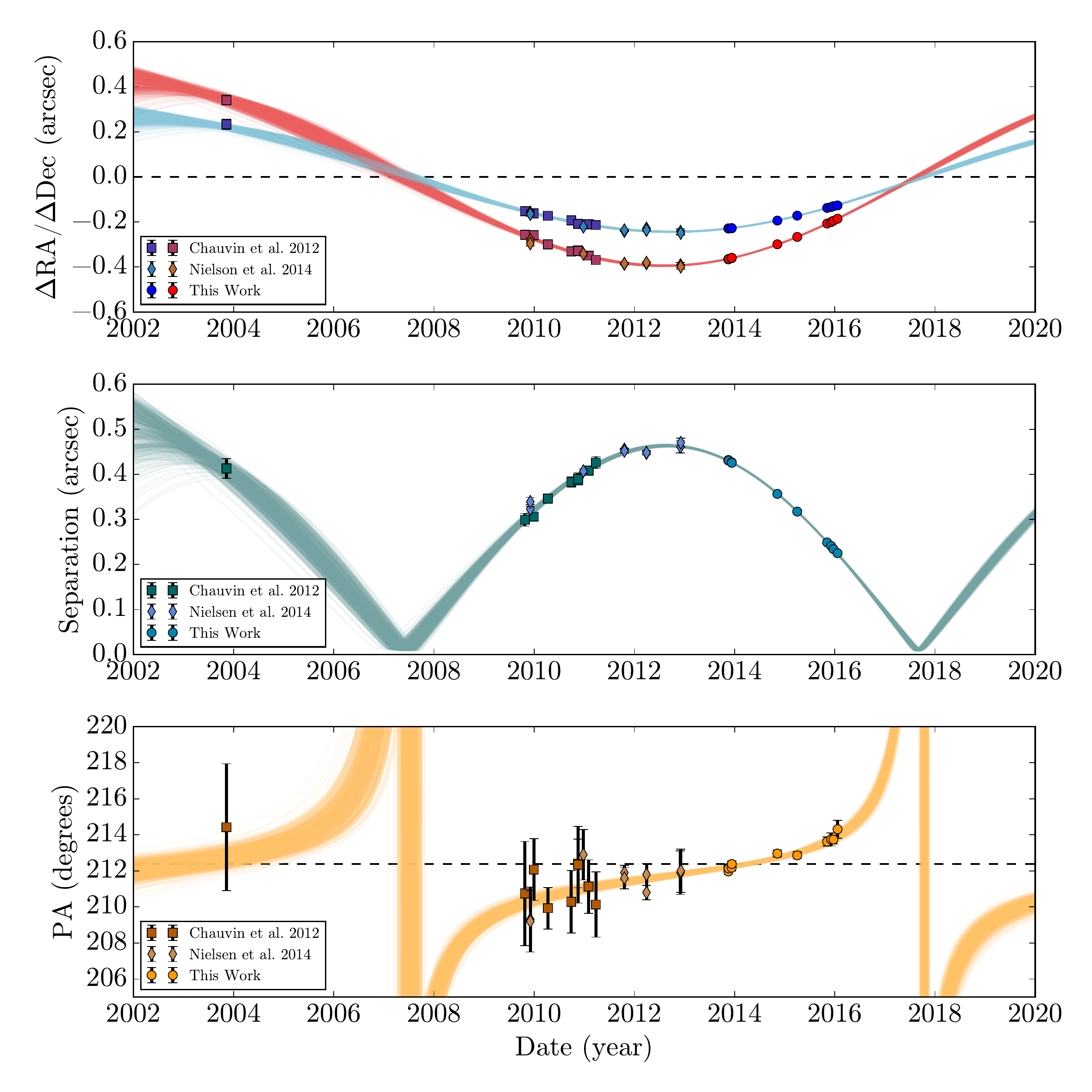}
\caption{\textit{(Top)} Offset of \betapicb{} in right ascension (blue) and declination (red) with respect to $\beta$ Pic as a function of time. We have plotted the measured data and 500 randomly-chosen accepted orbits from the MCMC sampler. \textit{(Middle)} Radial separation of \betapicb{} from the star as a function of time. The same 500 randomly-chosen orbits have are also plotted \textit{(Bottom)} PA as a function of time for the data and the 500 randomly-chosen orbits. To keep the data compact, we have wrapped PA by $180^{\circ}$ to only consider position angles between $180^{\circ}$ and $360^{\circ}$. This allows for easy comparison of the 2003 point, which is nominally at a PA of $34\fdg4$ but here displayed at a PA of $214\fdg4$. The dashed black line indicates a constant PA of $212\fdg4$, the weighted mean of all points. If the planet were to transit, we would not be able to see a significant deviation from constant PA in time. For all the plots, error bars are also plotted but many are too small to be seen on this scale. 
\label{fig:orbit-vs-time}}
\end{figure*}

The MCMC sampler was run for 30,000 steps using 1024 walkers at each of the 20 temperatures after 30,000 steps of ``burn-in'' to allow the walkers to converge. We thinned the chains to remove any remaining correlations, keeping every 75th step to result in effectively 400 samples per walker for a total of 409,600 samples to construct our posterior.

\begin{figure*}[!ht]
\plotone{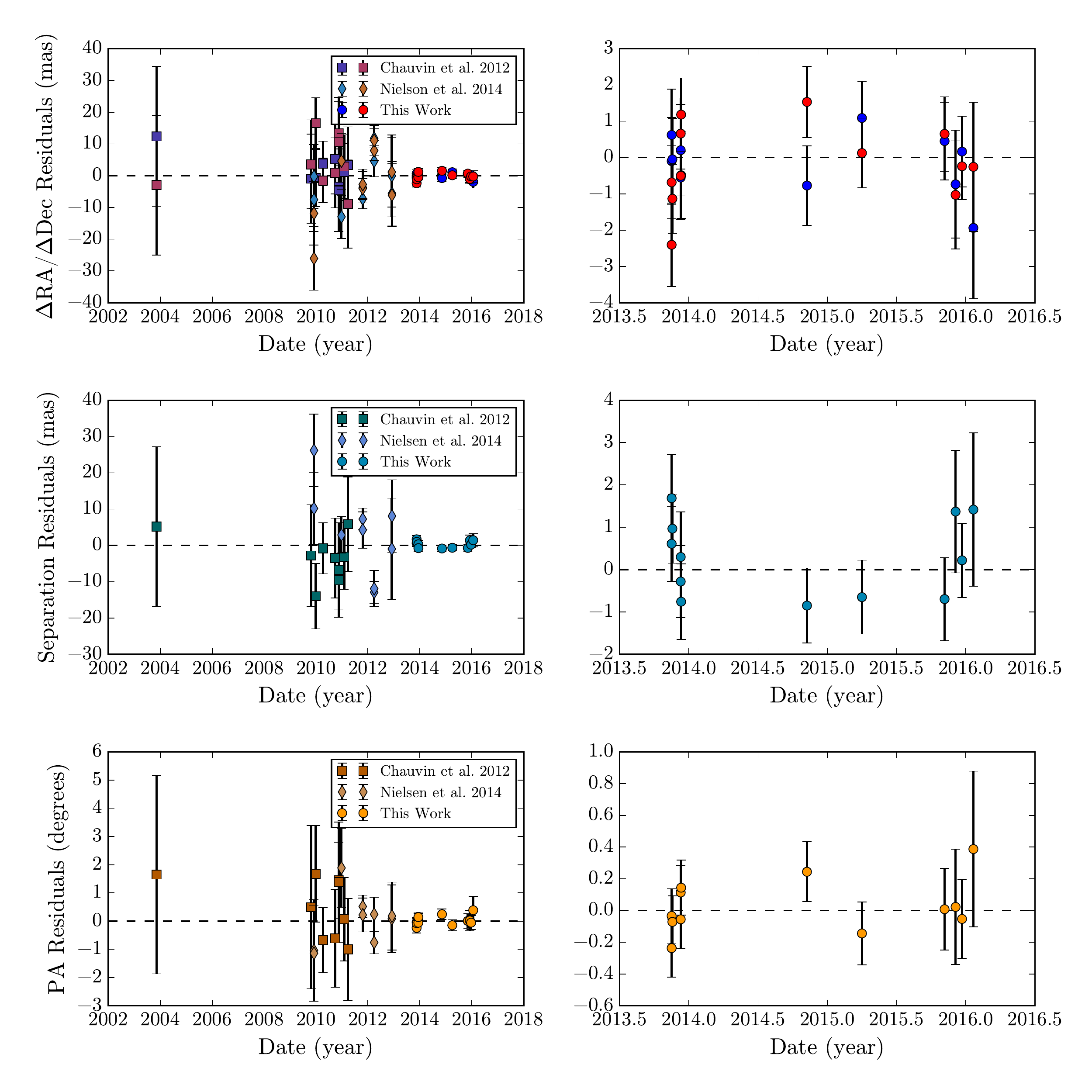}
\caption{Residuals to the orbit fit for the average of 500 randomly-chosen accepted orbits. The top row shows the residuals in $\Delta$RA (blue) and $\Delta$Dec (red) offset, which are the coordinates used in the MCMC analysis to fit the orbit. We also plot radial separation (middle row) and position angle (bottom row). The separation and position angle residuals were not optimized in the MCMC fit. The right column is a zoom-in of the left column showing only the residuals of the GPI astrometry. \label{fig:residuals}}
\end{figure*}

The posteriors on the seven parameters in our model are shown in Figure \ref{fig:orbit_triangle} and the 16\%, 50\%, and 84\% percentiles of the marginalized posterior distribution for each parameter is listed in Table \ref{tab:orb_params}. For comparison, we plot 500 randomly-chosen possible orbits along with the measured astrometry in Figure \ref{fig:orbit-vs-time}. The fit residuals of our measured GPI points are shown in Figure \ref{fig:residuals}. The residuals in RA and Dec offset (which are the parameters we use in our MCMC) are consistent with zero and do not show any systematics. The residuals in radial separation and position angle (PA), neither of which are used in our fit, are slightly further away from zero, but do not indicate any obvious errors in either our astrometry or our error estimates. 

To quantitatively assess the validity of our measurements, we use the reduced chi-squared ($\chi_{red}^2$) statistic, which measures how consistent our astrometry is with a Keplerian orbit. Ideally, $\chi_{red}^2$ would be unity if all measurements and uncertainties were accurate. However, biases and improperly estimated errors would cause it to deviate from unity. Due to systematic astrometric calibration offsets between instruments that have not been characterized, we expect  $\chi_{red}^2$ to be slightly above unity. For example, \citet{mmb15} reported a $\chi_{red}^2$ of $1.55^{+0.09}_{-0.05}$ when combining GPI astrometry, measured using standard techniques in the field, with previous astrometric measurements

Even with $\sim$2--4 times smaller error bars on the GPI points, the $\chi_{red}^2$ of our accepted orbits is $1.53^{+0.08}_{-0.06}$, unchanged from the $1.55^{+0.09}_{-0.05}$ reported in \citet{mmb15}. With BKA contributing twelve out of the thirty astrometry measurements, if the BKA technique contained biases larger than one milliarcsecond, they would have caused a significant increase in $\chi_{red}^2$. Likewise, if we had been too optimistic with our error estimates, $\chi_{red}^2$ should have also increased as the reported uncertainties would not have matched the scatter in the measurements. The lack of increase in $\chi_{red}^2$ indicates that the more precise astrometry from BKA is not biased and has accurate uncertainties.

In addition to the Keplerian orbit fit, we can examine the accuracy of our estimated one milliarcsecond uncertainties by looking at the astrometry measured on the same or consecutive days. As we do not expect the planet's position to change significantly in the span of a single day, the three measurements in November of 2013 and the three measurements in December of 2013 ought to be consistent with each other. Indeed, our measurements in Table \ref{table:astrometry} indicate that in both sets of astrometry, the measurements agree at the milliarcsecond level, confirming that our estimated uncertainties are accurate.

Thus, the well-behaved residuals of our GPI measurements and the lack of change in $\chi_{red}^2$ from \citet{mmb15} even with significantly smaller error bars lead us to conclude the measured astrometry using BKA are accurate and free from biases. The fact the residuals are consistent with zero, the lack of change in $\chi_{red}^2$, and the consistency of repeated measurements taken around the same time all indicate that the one milliarcsecond uncertainties estimated from BKA are also accurate. Together, these multiple assessments of the validity of BKA all indicate that this technique is both accurate and precise.

\section{Discussion}\label{sec:orbit}
\subsection{The Orbit of $\beta$ Pic b}
Having demonstrated the accuracy and precision of this new technique, we now analyze the new constraints on the orbit of \betapicb{}. Not surprisingly, the estimates for $a$ and $e$ have not changed significantly from \citet{mmb15} since the GPI points reanalyzed in this paper do not have a sufficiently long time baseline to constrain these parameters. As seen in Figure \ref{fig:orbit-vs-time}, all but one astrometric measurement is on one half of the orbit curve. The other half of the orbit is not as well constrained, leaving a degeneracy in $a$ and $e$. This degeneracy can be broken with more measurements obtained when the planet appears on the other side of the star. Better constraints on $a$ and $e$ will provide better insight on how \betapicb{} interacts with the debris disk and potential unseen planets in the system. The new total mass of the system, $M_T = 1.80^{+0.03}_{-0.04}$~$M_{\odot}$, is significantly higher than the $M_T = 1.61 \pm 0.05$~$M_{\odot}$ from \citet{mmb15}. This new total system mass, which effectively measures the mass of the star at this precision, is in better agreement with the stellar mass of $1.75$~$M_{\odot}$ derived from stellar photometry \citep{crifo97}.  Our measurement of the position angle of the ascending node, $\Omega$, slightly improves upon the value obtained by \citet{mmb15}, but the overall value remains consistent. Thus, \betapicb{} is still consistent with being the planet responsible for the known warp in the debris disk \citep[e.g.][]{dawson11}. Compared to \citet{mmb15}, the argument of the periastron, $\omega$ in our fit has increased from $156\degr^{+33}_{-76}$ to $206\degr^{+52}_{-13}$. This new value is consistent with $\omega=200\degr \pm 20\degr$ that is required for the falling evaporating bodies (FEBs) scenario proposed by \citet{Thebault2001} to explain redshifted absorption features in $\beta$ Pic’s spectrum. Note that under previous definitions of the orbital parameters, this has been expressed as $\omega=-70\degr \pm 20\degr$ from the line of sight. This scenario also requires a slightly eccentric orbit, which is consistent with our derived orbital parameters. For a more in-depth discussion of \betapicb's relationship to the debris disk and the FEB scenario, we direct the reader to \cite{mmb15}.

The biggest improvement in our understanding of the orbit of \betapicb{} is the improved constraint on the inclination of the orbit. We find the inclination to be $i = 88\fdg81^{+0.12}_{-0.11}$ which allows us to place the tightest constraints on the probability that \betapicb{} will transit its host star. Assuming an angular diameter of the star of 0.736~mas \citep{defrere12} and considering the range of $a$ from our orbit fit, we find that in most cases, we need $|i-90^{\circ}| < 0.05^{\circ}$ in order for the planet to transit. With our current constraints on the inclination, we have ruled out the possibility that \betapicb{} will transit at 10-$\sigma$ significance. This tight constraint on the inclination and transit probability is due to the slightly longer time baseline and the improved precision in the measured position angle of the GPI astrometry compared to \citet{mmb15}. For an edge-on orbit that transits the star, we should see no significant change in PA over time. However, Figure \ref{fig:orbit-vs-time} shows that the GPI points alone reveal a significant increase in PA over time. Thus, regardless of systematic astrometric calibration errors between instruments, we conclude that \betapicb{} will not transit its star.

\subsection{Hill Sphere Transit}
\begin{figure}[Ht]
\plotone{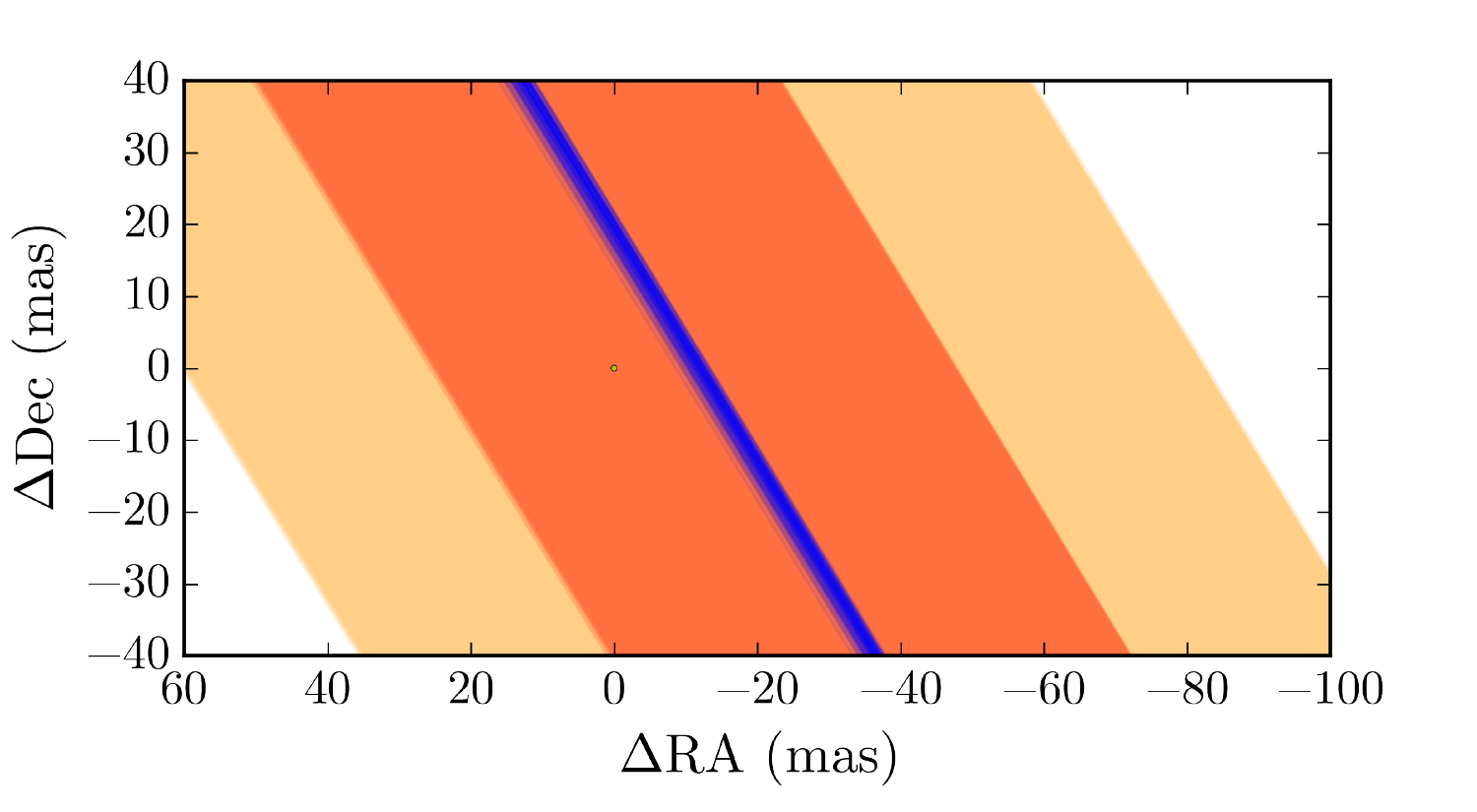}
\caption{The region in the sky that the Hill sphere of \betapicb{} will sweep across during the closest approach of the planet in 2017. 100 randomly-chosen accepted orbits (blue) are plotted along with the angular extents of their Hill spheres (light orange) and $1/2$ Hill spheres (dark orange). The star, $\beta$ Pic, is shown in its true angular size (small yellow dot). Both the Hill sphere and $1/2$ Hill sphere will pass in front of the star. Note that for clarity, we are not plotting the orbital path for \betapicb{} when it comes back in its orbit and passes behind the star. \label{fig:hillsphere_transit}}
\end{figure}

Unlike the planet, \betapicb{}'s Hill sphere, the region around the planet that could contain gravitationally bound circumplanetary material, will transit the star. We define the radius of the Hill sphere as $r_H \approx a (1-e)\sqrt[3]{\left(\frac{m}{3M}\right)}$ using the approximate form proposed in \citet{hamilton92} where $m$ is the mass of the planet and $M$ is the mass of the star. Assuming \betapicb{} follows a``hot-start" evolutionary track with a mass of $12.7 \pm 0.3$~$M_{\text{Jup}}$ \citep{morzinski15} and using the range of semi-major axes, eccentricities, and stellar masses from our MCMC orbit fit, $r_H = 1.165^{+0.013}_{-0.016}$~au ($59.9^{+0.7}_{-0.9}$~mas). 
Given that our prediction for the closest approach of \betapicb{} will be $9.9^{+0.9}_{-0.8}$~mas ($0.19 \pm 0.02$ projected au) from the star, the Hill sphere of \betapicb{} will transit the star, as shown in Figure \ref{fig:hillsphere_transit}.

From our MCMC orbit fit, we can compute  when the transit of the Hill sphere will occur. We pick five notable events to focus on: two are the ingress and egress of the Hill sphere which are the extrema in time between which any circumplanetary material could transit; another is the date of closest approach, which gives the opportunity to probe material closest in to the planet; and the last two are the transit of the sphere that is $1/2$~$r_H$ in radial extent. Almost all stable prograde circumplanetary orbits reside within $1/2$~$r_H$ \citep{shen08}, so it is more likely to find material within half a Hill sphere. In Table \ref{tab:orb_params}, we list our constraints on the date of these events, and in Figure \ref{fig:orbit_triangle}, we plot the posterior distribution for these events. The duration of the Hill-sphere transit will be long: ingress is $\sim$2017 April 3 and egress is  $\sim$2018 January 29. The 1/2 Hill sphere begins transit $\sim$2017 June 20 and ends transit on $\sim$2017 November 12, with closest approach on $\sim$2017 August 31. 

Given the RA of $\beta$ Pic ($05^h47^m17^s$), the star will not be visible from most ground-based observatories during almost the entire time between ingress of the Hill sphere and closest approach. Ground-based telescopes in Antarctica, airborne observatories capable of travelling to Antarctica, and space-based observatories would provide the only opportunities to observe $\beta$ Pic during this time period.  The second half of the Hill-sphere transit will be visible from most southern-hemisphere ground-based observatories, making $\beta$ Pic a well-suited candidate for photometric monitoring in late 2017. 

As it is a rare opportunity to probe circumplanetary material, it is not certain what will be seen when the Hill sphere of \betapicb{} transits the star. 
One possibility is that satellites could reside in the Hill sphere. To approximate the photometric transit depths, the 0.736~mas angular diameter of the star corresponds to 1.53 $R_\odot$ or 1.065$\times$10$^6$ km.  Moons as large as Ganymede ($r$ = 2630 km) or Io ($r$ = 1820 km) would give transit depths of 2.473$\times$10$^{-3}$ (2.7 mmag) and 1.711$\times$10$^{-3}$ (1.9 mmag), respectively. Detecting these photometric signatures will require a high cadence, as any single satellite orbiting \betapicb{} will have a transit duration of $\sim$2~days. Additionally, $\beta$ Pic is a variable star with pulsation timescales of $\sim$0.5~hours and variability amplitudes in \textit{B}-band of $< 5$~mmags \citep{koen03}, so careful modelling of stellar activity is necessary to be sensitive to these transit depths. 

Another possibility is that, as \betapicb{} is still young and evolving, it may harbor a circumplanetary disk or ring system comprised of leftover material from planet formation. Such a hypothesis is not unprecedented as \cite{kenworthy15} found evidence for a large circumplanetary disk around an unseen planet in the 1SWASP~J140747.93-394542.6 (hereafter J1407) system, which has a similar age \citep[$\sim$16 Myr;][]{mamajek12} and thus likely at a similar stage in its evolution. \cite{kenworthy15} interpreted the series of complex and deep eclipses in the J1407 lightcurve as from a circumplanetary disk 0.6~au in radial extent in the process of forming rings due to newly formed satellites. It is plausible that \betapicb{} can harbor a similar disk as a 0.6~au disk would be $1/2$ $r_H$ in extent and consistent with where we would expect stable orbits to reside around \betapicb{}. Additionally, since \betapicb{} is young, it is plausible that there is a large amount of circumplanetary material which has yet to be cleared out dynamically. Such a large disk would transit the star and be suitable for detection through photometric monitoring of the host star if the disk is inclined by $\gtrsim 18^{\circ}$ with respect to the orbital plane of \betapicb{}. To maintain such an inclination, the disk needs to be massive enough to prevent stellar tidal forces from aligning the disk to the orbital plane \citep{zanazzi16}.  

So far, there have not been many observational constraints of circumplanetary material around \betapicb{}. \citet{lecavelier97} reported a photometric event in 1981 and hypothesized it could be due to the transit of a planet that cleared out a hole in the debris disk around its Hill sphere. However, \citet{mmb15} show that the planet is not embedded in the debris disk and we definitely show the planet itself will not transit. Still, our orbit models give a $8$\% and $4$\% chance the photometric event coincided with the transit of the Hill sphere and 1/2 Hill sphere respectively, during which time material around the planet could have passed in front of the star. Additionally, a circumplanetary disk or ring may also be detectable through the planet's spectral energy distribution (SED). The planet's near-infrared spectrum would experience extinction if the dust resides between us and the planet, with the magnitude of extinction depending on the amount of dust. In the near infrared, the extinction would be greater at shorter wavelengths due to the increased scattering and absorption by dust and would produce a spectral slope in the planet's near-infrared SED. Dust around the planet would also scatter starlight, causing the planet’s SED to appear brighter in the optical, as has been postulated for Fomalhaut b \citep{kalas08}. Furthermore, the dust will produce millimeter emission. However, detecting circumplanetary material in the planet's SED will require being able to distinguish it from the circumstellar disk with precise spectral data.

\section{Conclusions}
In the first part of this work, we have presented a new technique for more precise and accurate astrometry of directly imaged exoplanets using a new analytical forward modelling approach in a robust statistical framework.
\begin{itemize}
\item Using the KLIP-FM framework presented in \citet{pueyo16}, we are able to analytically forward model the PSF of the planet through the data reduction process, giving us better information on the location of the planet. We apply KLIP-FM to GPI data on $\beta$~Pic and forward model the PSF of \betapicb{} using the open-source \texttt{pyKLIP} package.

\item For a close-in planet orbiting a bright star like in the case of \betapicb{}, we are limited by correlated speckle noise in our data. We developed a Bayesian framework utilizing Gaussian processes and MCMC to account for the correlated noise and to find the position of the planet simultaneously. 

\item With this technique, we have achieved the most precise astrometry on \betapicb{} to date. On most of our GPI datasets, we achieve $\sim$1~mas precision on the relative separation between \betapicb{} and its host star, a $\sim$2-4 fold improvement over previous techniques using the same data \citep{mmb15}. 

\item In datasets where the astrometry is limited by noise and not by astrometric calibration uncertainty, Bayesian KLIP-FM Astrometry approach should improve astrometric precision.
\end{itemize}

In the second part of this work, we apply our Bayesian KLIP-FM Astrometry technique to the orbit of \betapicb{}. 

\begin{itemize}

\item To validate this new astrometric technique, we used it to measure the position of \betapicb{} in twelve epochs of GPI data. We combined these twelve astrometric points with two previous astrometric monitoring campaigns and fit a Keplerian orbit using MCMC methods. We find the residuals to the fit are consistent with zero and show no apparent systematic trends, indicating that our fit is accurate and the uncertainties we estimate are reliable.

\item Due to the improved PA measurements from our technique, we have the tightest constraints on the inclination of the orbit and can exclude a possible transit of \betapicb{} at 10-$\sigma$ significance. 

\item While the planet will not transit, we are confident the Hill sphere around \betapicb{} will transit. The Hill sphere will begin transit at the start of April in 2017 and finish transiting at the end of January in 2018 with closest approach in the end of August in 2017. The transit of \betapicb{}'s Hill sphere should be our best chance in the near future to investigate young circumplanetary material. 

\end{itemize}

In the future, this MCMC forward modelling technique can be applied to photometry and spectral extraction alongside of astrometry of directly imaged exoplanets, allowing for improved characterization of their atmospheres. For \betapicb{}, continued monitoring of its orbit will yield more insight into the dynamics of the star system, although the planet will soon be too close to its star to be seen with current direct imaging instrumentation. However, once the planet appears on the other side, continued astrometric monitoring should be able to constrain the semi-major axis and eccentricity of the orbit much better, which will improve our understanding of how \betapicb{} perturbs the disk and if there are other planets perturbing \betapicb{}.

\acknowledgments
{\bf Acknowledgements:}
The GPI project has been supported by Gemini Observatory, which is operated by AURA, Inc., under a cooperative agreement with the NSF on behalf of the Gemini partnership: the NSF (USA), the National Research Council (Canada), CONICYT (Chile), the Australian Research Council (Australia), MCTI (Brazil) and MINCYT (Argentina).  We thank the National Science Foundation (NSF AST-1518332) and NASA (NNX15AC89G and NExSS program NNX15AD95G) for contributing to support of this research. Portions of this work were performed under the auspices of the U.S. Department of Energy by Lawrence Livermore National Laboratory under Contract DE-AC52-07NA27344. This research has made use of the SIMBAD database, operated at CDS, Strasbourg, France. The posterior distribution plots were made with \texttt{corner} \citep{corner16}. 

\appendix
\section{Analytic Forward Modelling of the Planet PSF after Stellar PSF Subtraction} \label{appendix:fmpsf}
Using the notation of \cite{pueyo16}, we can analytically forward model the PSF of the planet through the process of stellar PSF subtraction. Let us denote the target image as $T_{\lambda_p, t_p}(\mathbf{x})$, which is taken at wavelength $\lambda_{p}$ and time $t_p$ and contains the pixels $\mathbf{x}$ from which we want to subtract the stellar PSF.  

To forward model the PSF of the planet, we need a model of the PSF of the planet prior to stellar PSF subtraction. For our GPI data, we use the satellite spots to generate a realization of an unocculted and unprocessed point source 
as a function of wavelength. Using this model, we can generate ${A}_{\lambda_p,t_p}(\mathbf{x})$, a frame that consists solely of the unprocessed PSF of the planet in the target frame. The initial estimate for the planet's position in the frame is determined using an input separation and PA of the planet and the location of the star. Similarly, we can also generate an array of frames, each containing the unprocessed PSF of the planet for each image in the reference library, accounting for the fact that the planet position changes due to ADI field rotation and due to SDI rescaling to align speckles. In these frames, the PSF shape only depends on the wavelength of the frame and the position of the PSF depends on the apparent movement of the planet in the data due to ADI and SDI. This array of unprocessed PSFs of the planet will be used to calculate the perturbation of the KL modes due to the existence of the planet. 

Assuming a spectrum, $f$, of the planet in the case of IFS data (for imaging data, $f$ is just a scalar), we can compute $\Delta Z_k(\mathbf{x})$, the perturbation on the $k$th KL mode, ${Z}_k(\mathbf{x})$, due to the existence of a planet, using equation E18 or E20 of \citet{pueyo16}. We will not write out explicitly the exact formula to compute $\Delta Z_k$ here since we have not made any modifications to it. We note that we use Equation E18 from \citet{pueyo16} to compute $\Delta Z_k$ as it is computationally faster than Equation E20 for a fixed input spectrum.
We calculate the perturbed $\Delta Z_k$ for the first $k_{Klip}$ KL modes, where $k_{Klip}$ is the number of KL modes we choose to use in our PSF subtraction. 

From here, the forward-modeled PSF of the planet for this frame is computed as 
\begin{multline} \label{fmmath}
F_{\lambda_{p}, t_p}(\mathbf{x}) = {A}_{\lambda_p,t_p}(\mathbf{x}) \\
- \left[ \sum_{k=1}^{K_{Klip}} < {A}_{\lambda_p,t_p}(\mathbf{x}), Z_k (\mathbf{x}) > Z_k (\mathbf{x}) \right] \\
- \left[ \sum_{k=1}^{K_{Klip}} \left( <T_{\lambda_{p},t_p}(\mathbf{x}), Z_k (\mathbf{x})> \Delta Z_k (\mathbf{x})  \right. \right. \\
+ \left. \left. <T_{\lambda_{p},t_p}(\mathbf{x}), \Delta Z_k (\mathbf{x})> Z_k (\mathbf{x}) \right) \vphantom{\sum_{k=1}^{K_{Klip}}} \right],
\end{multline}
where $< \bullet , \bullet  >$ is the inner product.
Equation \ref{fmmath} is very similar to Eq. F7 in \citet{pueyo16}, but is focused on generating the forward model with a fixed input spectrum and not concerned with keeping the planet's spectrum as a free parameter for spectral extraction. As mentioned in \citet{pueyo16}, the term in the first square bracket is the over-subtraction term that is due to the projection of the KL modes on data with a planet in it, and the terms in the second square bracket are the self-subtraction terms due to the presence of a planet in the reference images influencing the KL modes. 

After generating forward models for each frame, we take the mean of all the forward models in time to give us a single forward modeled PSF cube, $F_{\lambda_{p}}(\mathbf{x})$ that still has spectral information in the third dimension. In principle, this PSF cube can be used to retrieve spectral information from the data, but it is outside of the scope of this paper. For our astrometry purposes, we also take the mean in wavelength, generating a single 2-D forward modeled PSF, which we call $F$, for each dataset.

\section{BKA Fit Residuals}\label{appendix:res}
\begin{figure*}
\begin{minipage}{0.5\textwidth}
    \centering
    \includegraphics[width=\textwidth, trim =36mm 30mm 15mm 30mm,  clip=true]{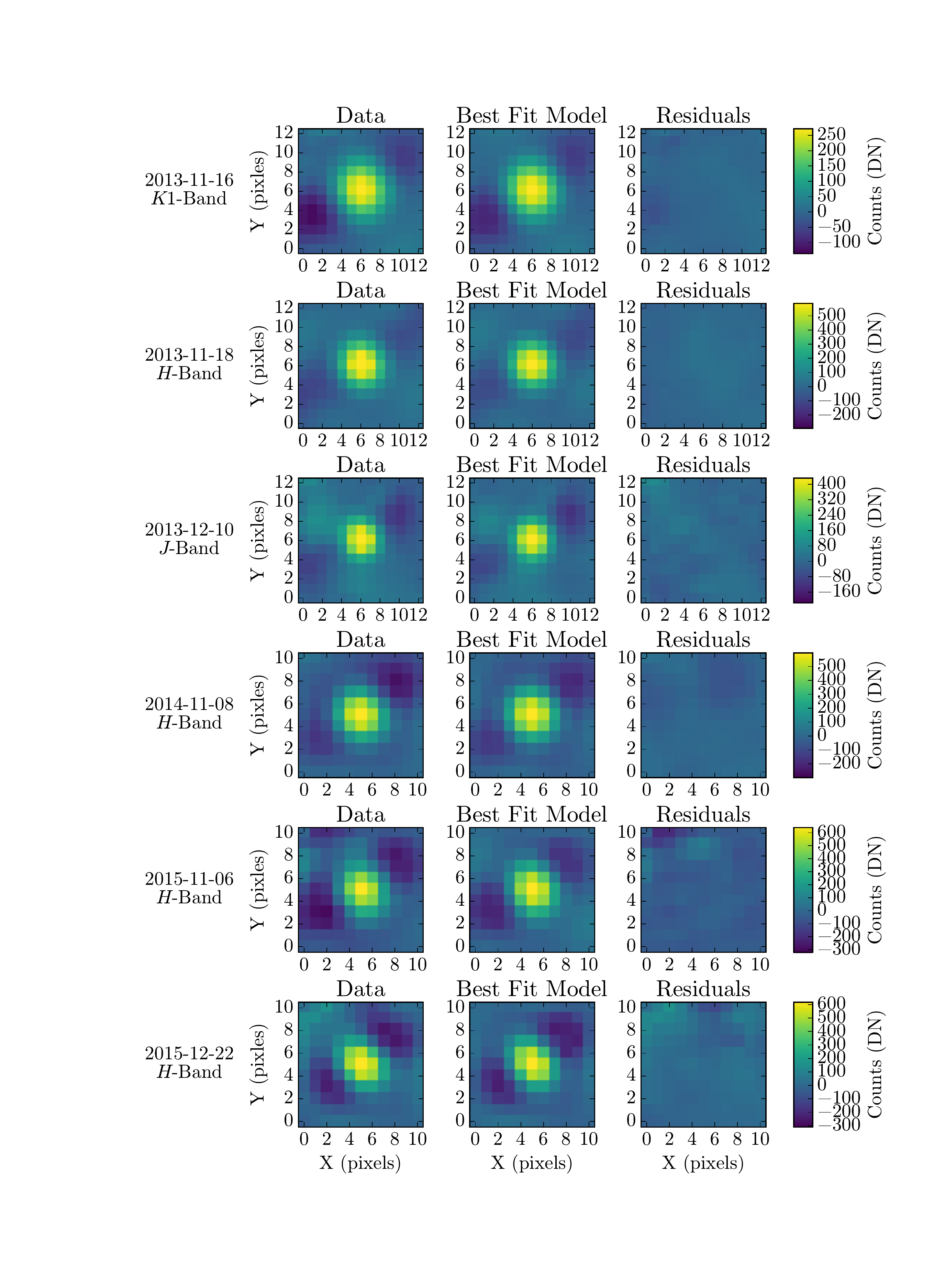}
\end{minipage}
\begin{minipage}{0.5\textwidth}
    \centering
    \includegraphics[width=\textwidth, trim = 36mm 30mm 15mm 30mm, clip=true]{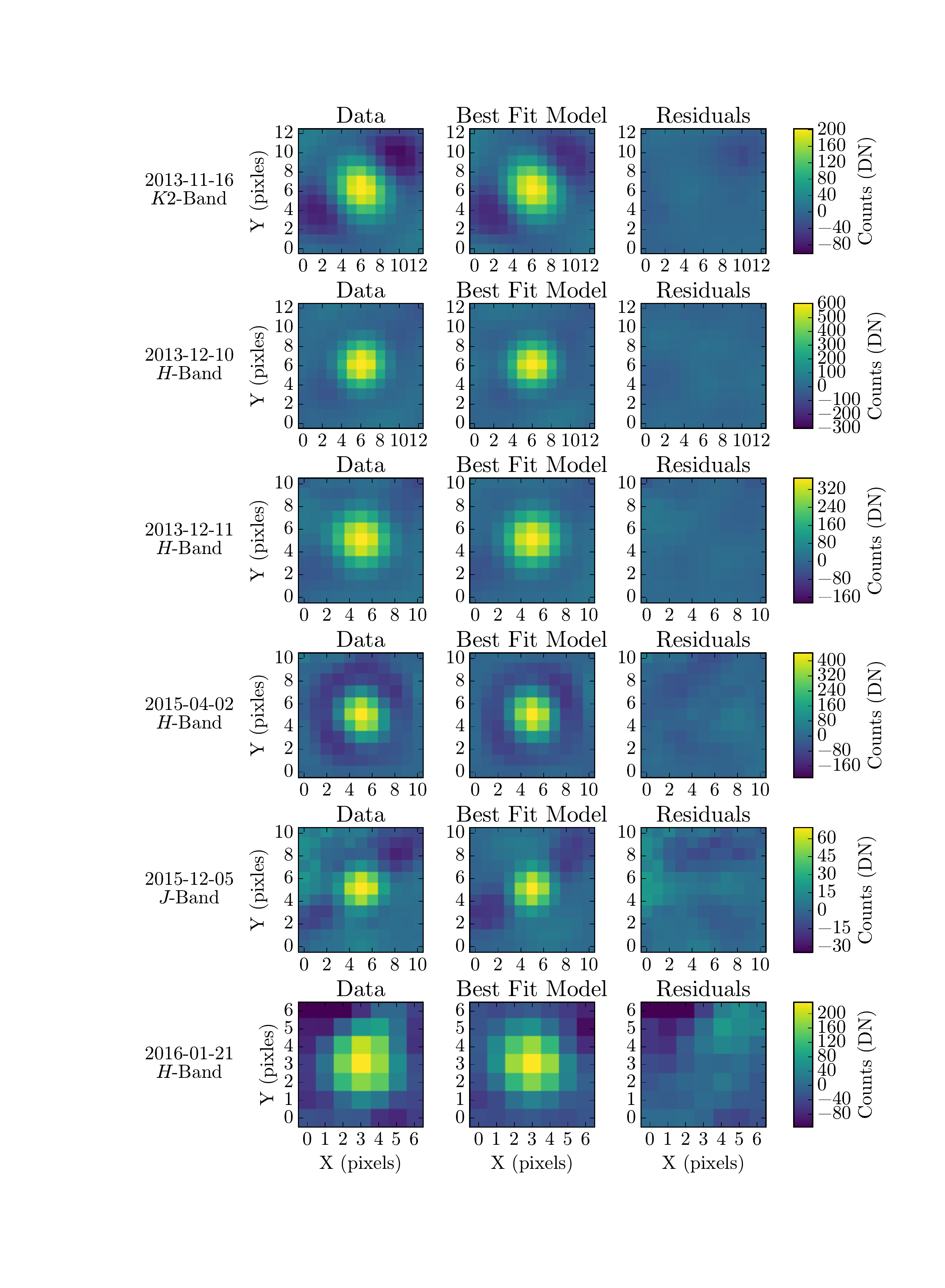}
\end{minipage}
\caption{The data, best fit forward model, and residual map after the model as been subtracted from the data for each of the twelve datasets. In each row, we plot two datasets. For each dataset, we plot the data (left), best fit forward model (center), and residual map (right) on the same color scale. While the scale of each dataset is different, zero is mapped to the same color for all the datasets. 
\label{fig:astrometry_residuals}}
\end{figure*}

In Figure \ref{fig:astrometry_residuals}, we plot the residuals to the best fit model for each epoch using BKA. As \betapicb{} moves closer to its host star, the magnitude of speckle noise increases relative to the signal of \betapicb{}. This is especially true in the last epoch when \betapicb{} was closest to its star. However, the fit was still accurate as the bright positive core of the PSF of the planet was successfully modeled and not seen in the residuals, which are consistent with noise.

\end{document}